# Towards smoother surfaces by applying subdivision to voxel data

A Michael Stock
University of Passau
Passau, Bavaria, 94032, Germany
+49 851 509 3149
stock@forwiss.uni-passau.de

Sergio López-Ureña
University of Valencia
Valencia, Province of Valencia, 46010, Spain

**Abstract**

In computed tomography, the approximation quality of a scan of a physical object is typically limited by the acquisition modalities, especially the hardware including X-ray detectors. To improve upon this, we experiment with a three-dimensional subdivision scheme to increase the resolution of the reconstructed voxel data.
Subdivision schemes are often used to refine two-dimensional manifolds (mostly meshes) leading to smoother surfaces. In this work, we apply a refinement scheme to three-dimensional data first, and only then, start the surface extraction process. Thus, the main subject of this work lies not on subdivision surfaces, but rather on subdivision volumes. In the volumetric case, each subdivision iteration consumes eight times more storage space than the previous one. Hence, we restrict ourselves to a single subdivision iteration. We evaluate the quality of the produced subdivision volumes using synthetic and industrial data. Furthermore, we consider manufacturing errors in the original and in the subdivision volumes, extract their surfaces, and compare the resulting meshes in critical regions. Observations show that our specific choice of a subdivision scheme produces smoothly interpolated data while also preserving edges.

## 1. Introduction

There are numerous different approaches to image super-resolution, especially in the two-dimensional case[1,2]. For three-dimensional data, there are slice-based and volume-based algorithms[3], as well as methods that rely on convolutional neural networks[4]. Another volume-based approach[5] smoothly uses second-order derivatives to preserve local average voxel values during the process.

Subdivision methods are especially well-known in the context of mesh refinement. An example for this is shown in Figure 1, where a smooth surface is produced as limit of a sequence of meshes. Research on subdivision surfaces has been done for decades[6] and it is still an ongoing subject[7].

In three dimensions, volume subdivision introduces seven new voxels at each iteration, see Figure 2, thus multiplying the required amount of storage space by a factor of eight. Therefore, only one subdivision step is considered.

In this work, a particular three-dimensional nonlinear subdivision scheme for isotropic voxel data upscaling is evaluated. This method is not only truly three-dimensional and deterministic, but also local due to the nature of the used subdivision scheme. Hence, tile-based processing of arbitrarily large input data is feasible. Also, no training or any kind of previous knowledge is required.

The proposed scheme is in parts inspired by theoretical work regarding a special family of non-oscillatory 6-point interpolatory subdivision schemes[8]. However, a different approach to avoiding oscillatory behaviour is chosen by using thresholded local second-order differences. Although some ideas between the one-[8], two-[9] and three-dimensional approaches can be shared, it should be emphasized, that it is more than just a three-dimensional extension of the aforementioned works.

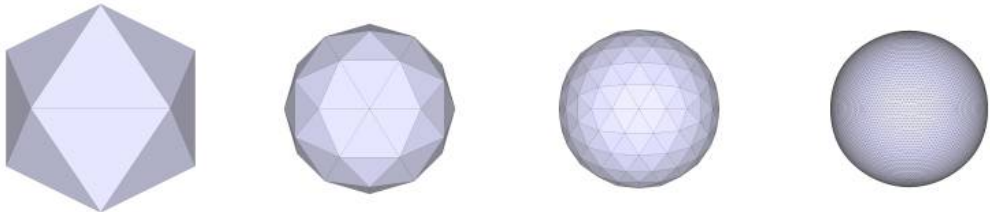

**Figure 1. Mesh subdivision process. A three-dimensional mesh is refined multiple times converging to a sphere**

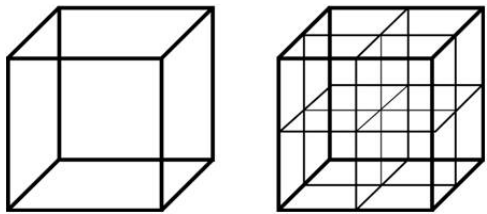

**Figure 2. Voxel subdivision step. To subdivide volumetric data on an isotropic grid of data points, each voxel is refined by doubling the resolution in each dimension**

## 2. Subdivision

The starting point is the observation that a fast method, which produces satisfying results in only one refinement iteration, is desired. This is especially urgent in the case for large three-dimensional data that is subject to current research[10].

### *2.1 A one-dimensional subdivision scheme*

Image and voxel data values usually represent local averages. Thus, for example, a subdivision scheme[8] in the *point-valued framework* might not be a methodologically good choice, even though it is centred, interpolatory and non-oscillatory. The 6-point scheme in the *cell-average framework* that produces polynomials of degree 5 is the following, which is the starting point in this work.

$$S(f)_{2i} = \frac{7}{512}f_{i-2} - \frac{63}{512}f_{i-1} + \frac{231}{256}f_i + \frac{69}{256}f_{i+1} - \frac{37}{512}f_{i+2} + \frac{5}{512}f_{i+3} \quad\ldots\ldots\ldots\ldots(1)$$

$$S(f)_{2i+1} = \frac{5}{512}f_{i-2} - \frac{37}{512}f_{i-1} + \frac{69}{256}f_i + \frac{231}{256}f_{i+1} - \frac{63}{512}f_{i+2} + \frac{7}{512}f_{i+3} \quad\ldots\ldots\ldots\ldots(2)$$

Now, the scheme is rewritten as sum of the 10th-order B-spline and second-order forward differences $\Delta^2 f_j = f_{j+2} - 2f_{j+1} + f_j$ as follows.

$$S(f)_{2i} = \frac{11}{1024}f_{i-2} + \frac{165}{1024}f_{i-1} + \frac{231}{512}f_i + \frac{165}{512}f_{i+1} + \frac{55}{1024}f_{i+2} + \frac{1}{1024}f_{i+3}$$
$$+ \frac{3}{1024}\Delta^2 f_{i-2} - \frac{285}{1024}\Delta^2 f_{i-1} - \frac{111}{1024}\Delta^2 f_i + \frac{9}{1024}\Delta^2 f_{i+1} \quad\ldots\ldots\ldots\ldots\ldots(3)$$

$$S(f)_{2i+1} = \frac{1}{1024}f_{i-2} + \frac{55}{1024}f_{i-1} + \frac{165}{512}f_i + \frac{231}{512}f_{i+1} + \frac{165}{1024}f_{i+2} + \frac{11}{1024}f_{i+3}$$
$$+ \frac{9}{1024}\Delta^2 f_{i-2} - \frac{111}{1024}\Delta^2 f_{i-1} - \frac{285}{1024}\Delta^2 f_i + \frac{3}{1024}\Delta^2 f_{i+1} \quad\ldots\ldots\ldots\ldots\ldots(4)$$

This can be reformulated by convolutions to

$$S(f)_{2i} = f \star a^0 = f \star b^0 + \Delta^2 f \star c^0 \quad\ldots\ldots\ldots\ldots\ldots\ldots\ldots\ldots\ldots\ldots(5)$$

$$S(f)_{2i+1} = f \star a^1 = f \star b^1 + \Delta^2 f \star c^1 \quad\ldots\ldots\ldots\ldots\ldots\ldots\ldots\ldots\ldots\ldots(6)$$

Note that the coefficients of $a^0, b^0, c^0$ are the same as in $a^1, b^1, c^1$ but in reversed order in this case. One can show that this scheme approximates cell averages of smooth functions with order 6, i.e., one can calculate the following for $h > 0$, a smooth function $F$, and cell averages $f_i = \int_{[i-h/2, i+h/2]} F(t)dt$.

$$S(f)_{2i} = \int_{[i-h/2, i]} F(t)dt + O(h^6) \quad\ldots\ldots\ldots\ldots\ldots\ldots\ldots\ldots(7)$$
$$S(f)_{2i+1} = \int_{[i, i+h/2]} F(t)dt + O(h^6) \quad\ldots\ldots\ldots\ldots\ldots\ldots\ldots\ldots(8)$$

In this work, applications of this scheme are considered, whereas further detailed mathematical analysis is still ongoing.

*2.2 Treating isolated discontinuities*

To avoid the Gibbs-like phenomenon, i.e., unwanted jumps in resulting subdivided data, the following function is introduced when analysing a function $f$.

$$\phi(\Delta^2 f_i, K_i) = \text{sign}(\Delta^2 f_i) \min(|\Delta^2 f_i|, K_i) \dots\dots\dots\dots\dots\dots\dots\dots\dots(9)$$

where

$$K_i = \tau \min\{|\Delta f_{i+j}| + |\Delta f_{i+j+1}|, j = -2, -1, 0, 1, 2\} \dots\dots\dots\dots\dots(10)$$

provides bounds for the second-order differences depending on the *tension parameter* $\tau > 1$, thus making the scheme nonlinear and non-oscillatory. Note that this holds due to $|\Delta^2 f_{i+j}| \leq |\Delta f_{i+j}| + |\Delta f_{i+j+1}|$, using the first-order differences $\Delta f_j = f_{j+1} - f_j$, for large enough $\tau$, which is thus a globally data-dependent parameter, and also note that $K_i$ and $\phi$ are functions that depend on local smoothness of the data around the position $i$. Thus, in smooth regions, the scheme behaves like the cell-averaging scheme, while near discontinuities, it behaves like the B-spline, resulting in smoothly shaped discontinuities.

*2.3 Building a non-oscillatory three-dimensional subdivision scheme*

In three dimensions, we apply the tensor product of the $10^{\text{th}}$-order B-spline-based scheme and afterwards, a trivariate extension of the bounding functions above is applied. Instead of the technical details, only the intuition is provided here. As the schemes from Section 2.1 can be decomposed into two parts, one of them being second-order differences, there are four types of convolution operations in three dimensions that include none, one, two, or three of the three directional second-order differences, that are added together. For each summand, depending on the number and direction of these differences, a corresponding threshold function must be applied that, just as above, depends on local sums of the first-order differences. Thus, the nonlinearity is a consequence of data-dependent refinement to avoid singularities. The method itself is isotropic due to the fixed grid-based position of voxels in the three-dimensional space, and it preserves local average voxel values.

*2.4 The algorithm in three dimensions*

Building the algorithm is split into two steps: first, the linear method without thresholding of the second-order differences is defined. Then, a three-dimensional version of the nonlinear filter $\phi$ from Section 2.2 is added.

Let $g$ be three-dimensional data, and $a, b, c$ as in the formulas (5) and (6). In the following, the dimensional application will be indicated by subscripts $x, y, z$, whereas the superscripts indicate the coefficients depending on the position of newly generated data in Section 2.1. Here, there are now eight different positions in three dimensions since every data point is transformed into eight new ones, whereas in one dimension the original data point is replaced by two. The second-order differences $\Delta_x^2 g, \Delta_y^2 g$, and $\Delta_z^2 g$, are, for $w \in \{x, y, z\}$, replaced by the discrete convolution with the vector $d_w^2$ in the expression $\Delta_w^2 g = (1, -2, 1)_w \star g = d_w^2 \star g$.

Now, based at a three-dimensional data position $p$, the subdivision-generated data at positions $2p + (k, l, m)$, i.e., $S(g)_{2p+(k,l,m)}$, where $(k, l, m) \in \{0,1\}^3$, is to be calculated.

*2.4.1 A linear volumetric subdivision algorithm*

In three dimensions, the linear algorithm with unmodified second-order differences is as follows.

$$\begin{aligned} S(g)_{2p+(k,l,m)} &= \left(g \star a_x^k \star a_y^l \star a_z^m\right)_p \\ &= \left(g \star b_x^k \star b_y^l \star b_z^m\right)_p \\ &+ \left(g \star b_x^k \star b_y^l \star d_z^2 \star c_z^m\right)_p + \left(g \star b_x^k \star d_y^2 \star c_y^l \star b_z^m\right)_p \\ &+ \left(g \star d_x^2 \star c_x^k \star b_y^l \star b_z^m\right)_p + \left(g \star b_x^k \star d_y^2 \star c_y^l \star d_z^2 \star c_z^m\right)_p \\ &+ \left(g \star d_x^2 \star c_x^k \star b_y^l \star d_z^2 \star c_z^m\right)_p + \left(g \star d_x^2 \star c_x^k \star d_y^2 \star c_y^l \star b_z^m\right)_p \\ &+ \left(g \star d_x^2 \star c_x^k \star d_y^2 \star c_y^l \star d_z^2 \star c_z^m\right)_p \end{aligned} \quad (11)$$

Due to symmetry and dimensional independence of the filters, the convolutions commute. Since the second-order differences occur multiple times, they may be cached for a fast implementation. However, the corresponding memory footprint must be considered.

*2.4.2 A nonlinear, non-oscillatory subdivision algorithm*

Using commutativity, second-order expressions containing convolutions of $g$ with $d_x^2, d_y^2, d_z^2$, i.e., terms like $\Delta_x^2 g, \Delta_x^2 \Delta_y^2 g, \Delta_x^2 \Delta_y^2 \Delta_z^2 g$, are evaluated first before completing the summand calculations. As in Section 2.2, a three-dimensional version of the filter $\phi$ is applied. The resulting terms replace the second-order differences which are then further processed as in the linear scheme above. This yields a nonlinear scheme that dampens oscillations: for example, at position $p$, the expression $\Delta_y^2 \Delta_z^2 g_p$ is replaced by $\phi(\Delta_y^2 \Delta_z^2 g_p, K_p^{y,z})$ where $K_p^{y,z}$ depends on second-order differences in directions $y, z$. This is done for all summands containing any number of second-order differences. The exact calculation of $K_p^{y,z}$ and the other thresholds at position $p$ is too technical for the scope of this work.

## 3. Experiments

In this section, one synthetic dataset and three instances of industrial CT data are processed, and the algorithm is analysed in terms of the peak signal-to-noise ratio (PSNR), total variation (TV) norm and other standard metrics, visual performance, and dimensional measurements.

*3.1 A synthetic example*

First, a synthetic dataset of size (256, 256, 256) with 32-bit floating-point voxels containing thin, flat, cubic, and two spherical structures, is examined. The voxel value range is between zero and one. Except for one sphere in this dataset, all objects take the value one. The other sphere has linearly decreasing values: one in the centre, zero at the surface.
Figure 3 shows that oscillations from the theoretical considerations above appear already in this very simple dataset: under- and overshoots can be observed at the border of the planes and spheres. After applying the linear scheme, the voxel value range is increased to the interval $[-0.31, 1.5]$, which is almost twice as big as the original range. The non-oscillatory, nonlinear scheme produces a dataset with values in the interval $[0,1]$ as desired. Note that these numbers are not rounded: the smallest value is indeed zero, the largest indeed one (within machine precision). Furthermore, it is observed that the inner part of the spheres is identical (and exact) for both schemes. Their differences are found in the boundary, where the linear scheme exhibits under- and overshoots.

### 3.2 Subsampled data analysis

Subsampling input data with subsequent upsampling[5] gives a first quantitative evaluation of the subdivision method on real data. Here, the CT scan of a Mahle motor piston is used, where the aluminium piston body, the iron ring, the fixation, and the surrounding air can be identified, see Figure 4. The piston is stored as a volume of dimensions (464, 464, 414) filled with unsigned 16-bit integers representing voxels with an edge length of 330 µm. Table 1 shows that L2-based and voxelwise metrics are negatively impacted by the usage of the nonlinear scheme in contrast to the TV norm, which is expected due to the increased smoothness that is caused by bounding the second-order differences. The slice view of the voxelwise absolute differences in Figure 5 demonstrates how the linear scheme is closer to the original data compared to the nonlinear one. The biggest differences are visible where the material changes, which is also a direct result of bounding the second-order differences.

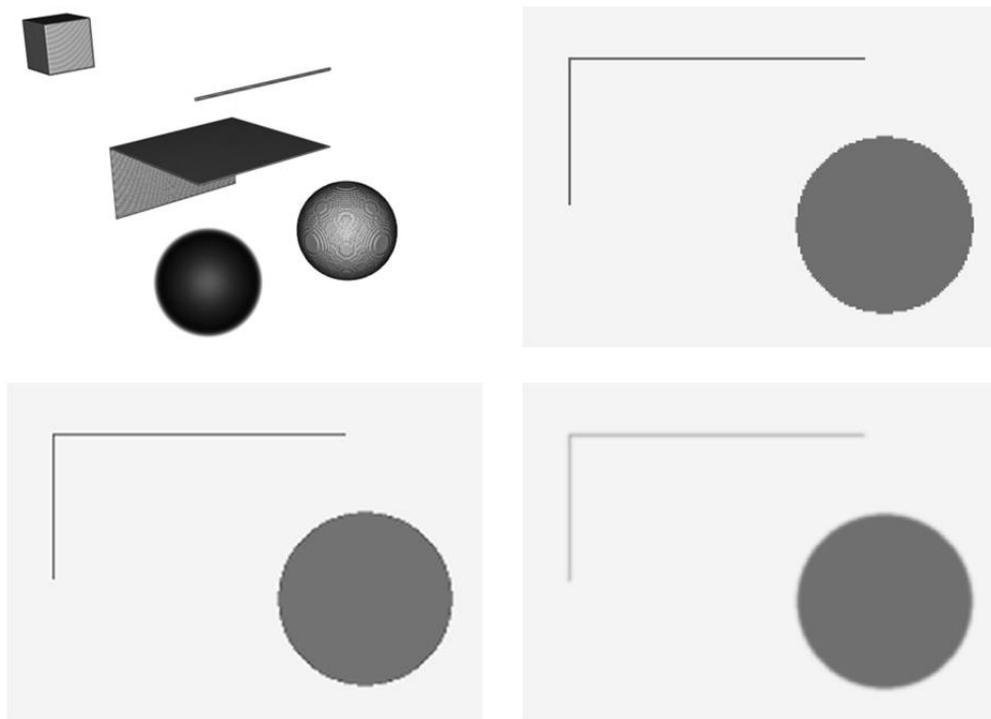

**Figure 3. Top row, left: three-dimensional view of the phantom containing a cube, a stick, two attached planes, one solid sphere, and one sphere with decreasing values towards the surface.**
**Top row, right, and bottom row, left and right: slices through regions of interest of the original data and the linear and nonlinear ($\tau = 2$) schemes, respectively, showing that discontinuities lead to oscillations in the linear case.**
**Note that the maximal value of $1.5$ is mapped to dark grey, and the minimal value of $-0.31$ to white, while zero is mapped to light grey. The results are similar for the shapes that are not shown here**

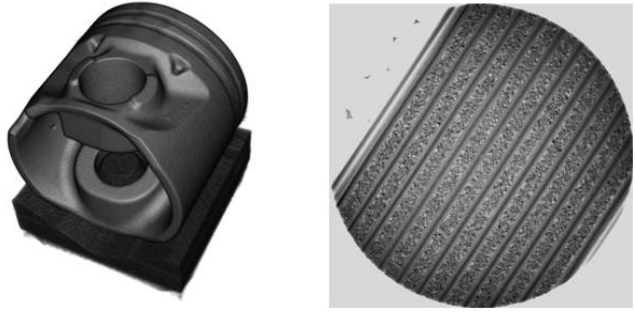

**Figure 4. A 3D view of the Mahle motor piston and a slice view of the flat battery dataset. The values are logarithmised for more contrast. The right slice is 4.6 mm wide in both dimensions**

**Table 1. Motor piston metrics for the linear and nonlinear schemes**

|  | Linear scheme | Nonlinear scheme ($\tau = 2$) |
|---|---|---|
| PSNR | **49.5** | 44.1 |
| Relative L2 error | **4.93 %** | 9.10 % |
| Median relative voxelwise error | **0.137 %** | 0.165 % |
| 99 % quantile of relative voxelwise error | **1.12 %** | 2.41 % |
| Relative TV norm | 82.4 % | **64.3 %** |

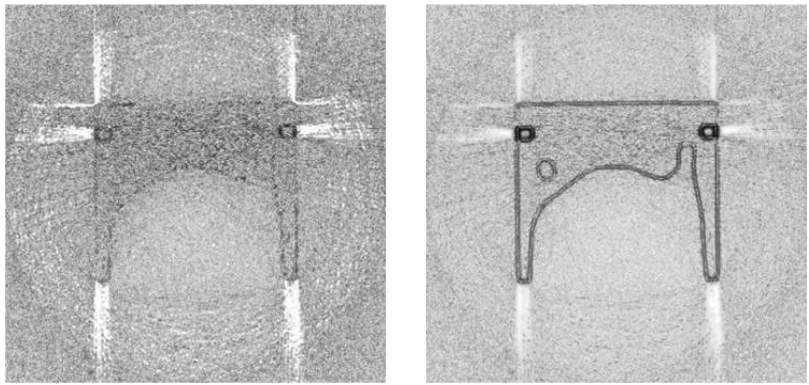

**Figure 5. Slice views of voxelwise absolute differences between the original and the subsampled and upsampled piston data for the linear and nonlinear ($\tau = 2$) scheme. Gamma correction was applied for increased contrast**

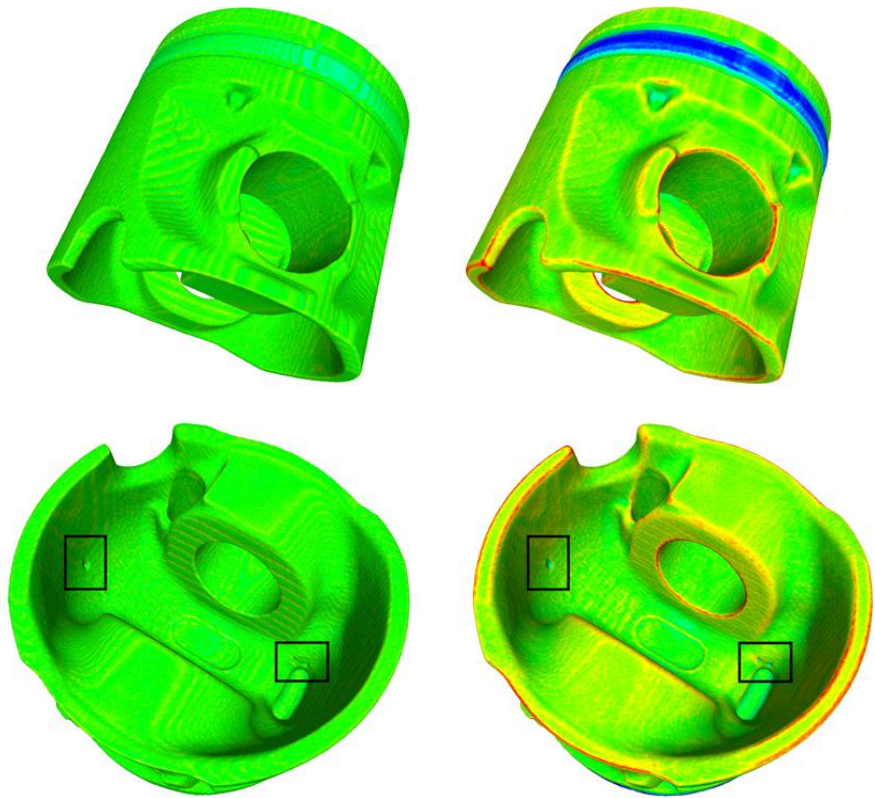

**Figure 6.** Left: mesh distance for the linear scheme, right: mesh distance for the nonlinear scheme ($\tau = 2$) using the same color scale. Bottom row: highlighted manufacturing errors, visible using both schemes

Finally, the focus is now changed to isosurface meshes extracted from subsampled and upsampled volume data. Figure 6 shows that the linear scheme retains sharp edges best. Still, manufacturing errors are also visible in the case of the nonlinear scheme.

*3.3 Improving compressed data for visualization*

Lossy compression methods may decrease the overall image quality, especially if the given data is very inhomogeneous[11]. A CT scan of a flat battery pouch, which was provided by the European Synchrotron Radiation Facility in Grenoble, see Figure 4, shows the granular structure at a resolution of 2.24 µm, which is very challenging for compression algorithms that are mainly designed for piecewise homogeneous data. Its dimensions are (2048, 2048, 2048) and it contains unsigned 16-bit integer values and thus has a size of 16.0 gigabytes. The wavelet-based compression[11] created a file of 1.31 gigabytes, thus reaching a compression factor around 12.2. Applying the subdivision methods to a region of interest of size (250, 250, 250) reduces artifacts from strong compression, especially when the nonlinear scheme is used, as Figure 7 shows. It took a computer with eight cores at 2.9 GHz around 13 seconds for the linear scheme and 14 seconds for the nonlinear scheme. Note that no bounding function $\phi$ is evaluated in the linear case. Table 2 compares the subdivided compressed volumes to the original subdivided volumes with the best scores after nonlinear-subdivision-based processing with respect to all tested metrics since a large PSNR, small L2 and voxelwise errors, and a small TV norm are desired. This demonstrates how the nonlinear scheme can improve the image quality both quantitatively and qualitatively.

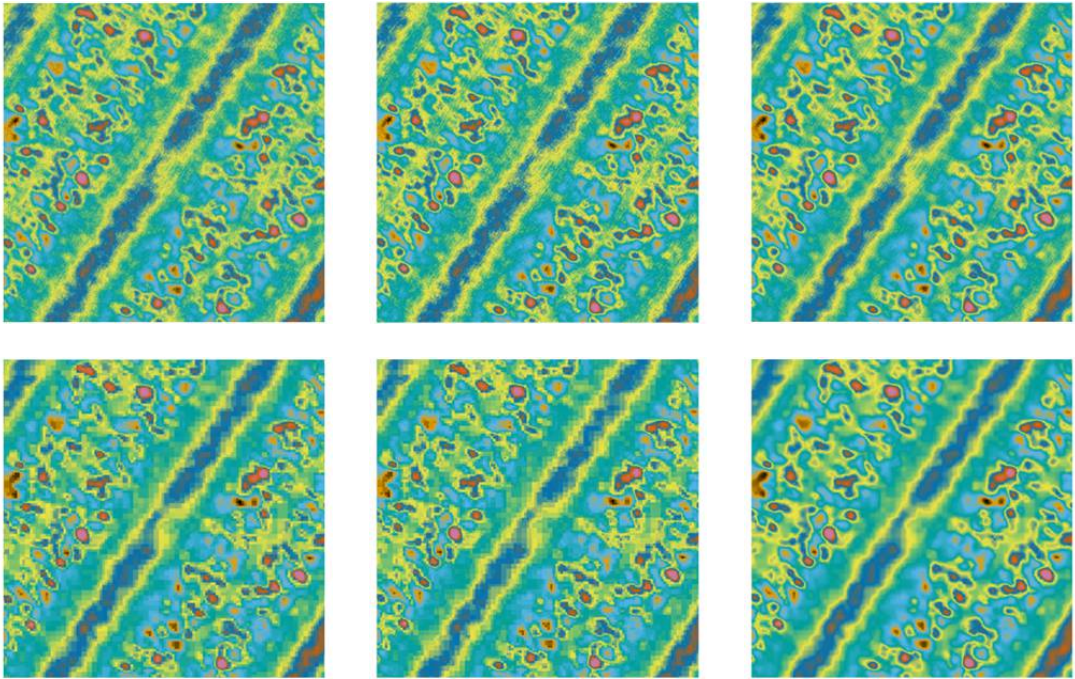

**Figure 7. Cross-sections of regions of interest of the flat battery dataset. Every slice corresponds to a region that is 0.56 mm wide in both dimensions.
Top row: slice views of the original data, after linear subdivision and after nonlinear subdivision ($\tau = 2$) using the original data,
bottom row: slice views of the compressed data, after linear subdivision and after nonlinear subdivision ($\tau = 2$) using the compressed data**

**Table 2. Flat battery region metrics. The original data is compared with the compressed data, and the linear and nonlinear schemes using original and compressed data for both schemes**

|  | Compressed vs. original | Linear scheme | Nonlinear scheme ($\tau = 2$) |
|---|---|---|---|
| PSNR | 43.1 | 44.3 | **48.5** |
| Relative L2 error | 23.5 % | 20.4 % | **12.9 %** |
| Median relative voxelwise error | 1.33 % | 1.15 % | **0.74 %** |
| 99 % quantile of relative voxelwise error | 5.51 % | 4.75 % | **2.95 %** |
| Relative TV norm | 92.1 % | 91.7 % | **82.3 %** |

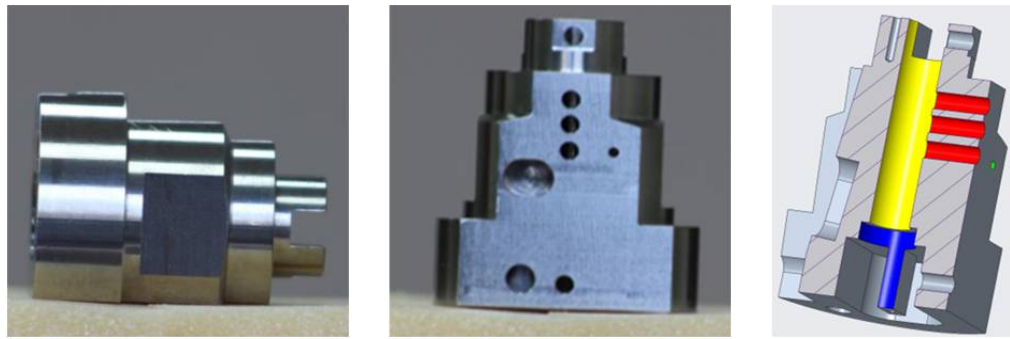

**Figure 8. From left to right: the aluminium test specimen in two different positions, and a visualization of the six measurement areas (blue, yellow, red, and green) for diameters and roundness of the test specimen**

*3.5 Metrological experiments*

In this section, the focus lies on data from a particular calibrated object for which there is exact geometric information. In this analysis, the aluminium test specimen data from our own previous work[11] is used, detailed experimental set-up can be found there. The object was scanned and reconstructed for 20 times and the data is stored in the unsigned 16-bit integer format with a voxel size of 50.03 μm and dimensions of (992, 992, 850). For every scan, the set of 25 features is processed before and after subdividing each of the scans using the nonlinear method with a tension parameter of $\tau = 2$. For the subdivision step, the same central region with voxel dimensions of (512, 512, 512) in each reconstruction is selected resulting in a subdivision volume with dimensions (1024, 1024, 1024). The measured diameters and distances from the original and subdivided data are compared.

*3.5.1 Metrological aspects of the aluminium test specimen*
Figure 8 shows the aluminium test specimen and six measurement regions where diameters are examined using the WinWerth metrology software. In total, eight feature regions are considered in terms of roundness, diameter, distance, perpendicularity, and concentricity. For brevity, only diameters, distances and roundness are considered in detail, whereas the remaining features will be considered only briefly.

*3.5.2 Deviations from calibrated measurements*
The deviations of original and subdivided volumes from calibrated measurements are compared for 120 diameters, 120 roundness measurements, and 60 distances, 6, 6 and 3 for every scan respectively, see Table 3. The results indicate that accuracy suffers especially for the distances under the proposed nonlinear subdivision scheme, whereas the diameters and roundness results are more stable and consistent with previous measurements[11]. Regarding the other features, it is expected that perpendicularity suffers most, more than doubling the average deviation from 26.9 μm from the calibrated measurement to 64.8 μm, due to the smoothing effect of the nonlinear subdivision scheme. However, the concentricity feature improved from an average deviation of 19.8 μm to 5.11 μm indicating that the cylindrical shape of that region was upscaled well.

Table 3. Dimensional metrology deviations of the aluminium specimen described in Section 3.5 compared with the calibrated measurements for the original and subdivided (nonlinear, $\tau = 2$) data

|  | Average deviation (μm) | Standard deviation (μm) | Maximum deviation (μm) | 0.95 quantile (μm) |
|---|---|---|---|---|
| Diameters (original) | **7.91** | **4.07** | **21.7** | **14.9** |
| Diameters (subdivided) | 8.07 | 8.17 | 74.9 | 14.2 |
| Roundness (original) | **23.3** | **11.6** | **48.1** | **43.1** |
| Roundness (subdivided) | 28.94 | 14.5 | 58.6 | 46.8 |
| Distances (original) | **5.90** | **3.58** | **12.0** | **10.9** |
| Distances (subdivided) | 10.1 | 6.50 | 25.9 | 21.9 |

# 4. Conclusions

This work shows that subdivision-based techniques for enhancing volumetric data are not only a tool that provides visually appealing results, but the method shows expected and satisfying results with regards to some, but not all metrological aspects. Although the procedure was restricted to only one iteration due to the high computational effort, it can help to visually compensate certain compression artifacts as well. This is particularly useful in the context of CT data if high-resolution images are desired. Except for the tension parameter, all calculations, i.e., the convolutional operations and the modification of the second-order differences, are local, yielding a method that has no limit with regards to the size of the input data, and that can be implemented in a data-parallel fashion.

**Acknowledgments**

This work was supported by the German Federal Ministry of Education and Research (BMBF) via the joint project 'BM18: High resolution industrial tomography beamline for large objects' under the title 05E2019, in collaboration with the institute FORWISS, University of Passau, and the Fraunhofer Gesellschaft zur Förderung der angewandten Forschung e.V., and the European Synchrotron Radiation Facility in Grenoble. We also thank Severin Nusser (Deggendorf Institute of Technology) for providing the measurement data in Section 3.5.
The second author has been supported by grant MTM2017-83942 funded by Spanish MINECO, by grant PID2020-117211GB-I00235 funded by MCIN/AEI/10.13039/501100011033 and by grants FPU14/02216 and EST17/00546 funded by Spanish MCIU.

# References

1. K Li, S Yang, R Dong, X Wang, J Huang, 'A Survey of Single Image Super Resolution Reconstruction', IET Image Processing, Vol 14, September 2020.
2. Wang Z, Chen J, Hoi, S C. 'Deep learning for image super-resolution: A survey.' IEEE transactions on pattern analysis and machine intelligence, Vol 43, No 10, pp 3365-3387, 2020
3. Q Wang, Y Tao, C Wang, F Dong, H Lin, G Clapworthy, 'Volume Upscaling Using Local Self-Examples for High Quality Volume Visualization', 2013 International Conference on Computer-Aided Design and Computer Graphics, pp 298-305, 2013.
4. Z Zhou, Y Hou, Q Wang, G Chen, J Lu, Y Tao, H Lin, 'Volume Upscaling with Convolutional Neural Networks', Proceedings of the Computer Graphics International Conference, No 38, June 2017.
5. A Giachetti, J A I Guitián, E Gobbetti, 'Edge Adaptive and Energy Preserving Volume Upscaling for High Quality Volume Rendering', Eurographics Italian Chapter Conference, Vol 10, pp 17-23, 2010.
6. M Sabin, 'Recent progress in subdivision: a survey', Advances in multiresolution for geometric modelling, pp 203-230, 2005.
7. T J Cashman, 'Beyond Catmull–Clark? A survey of advances in subdivision surface methods', Computer Graphics Forum, Vol 31, No 1 , 2012.
8. R Donat, S López-Ureña, M Santágueda, 'A family of non-oscillatory 6-point interpolatory subdivision schemes', Advances in Computational Mathematics, Vol 43, No 4, pp 849-883, 2017.
9. C Conti, S López-Ureña, 'Non-oscillatory butterfly-type interpolation on triangular meshes', Submitted.
10. S Kasperl, R Hanke, S Oeckl, P Schmitt, N Uhlmann, D Heinz, G Herl, J Hiller, A Kämmler, T Miller, A M Stock, T Sauer, 'Digitalisierung, Verarbeitung und Analyse kultureller und industrieller Objekte: Wertschöpfung aus großen Datenmengen', NDT.net Issue, Vol 23, No 9, September 2018.
11. A M Stock, G Herl, T Sauer, J Hiller, 'Edge-preserving compression of CT scans using wavelets', Insight, Vol 62, No 6, pp 345-351, June 2020.